\documentstyle[12pt,epsf]{article}

\makeatletter
\@addtoreset{equation}{section}
\@addtoreset{footnote}{page}
\makeatother

\newcommand{\ol}{\overline}
\newcommand{\del}{\partial}

\newcommand{\gym}{g_{Y\!M}}

\def\matt[#1,#2,#3,#4]{\left(%
\begin{array}{cc} #1 & #2 \\ #3 & #4 \end{array} \right)}

\def\v2#1{\vv2[#1]}
\def\vv2[#1,#2]{\left(%
{#1 \atop #2}\right)}

\begin{document}

\begin{titlepage}
\vspace*{-2.0cm}
\null
\begin{flushright}
hep-th/0404205 \\
YITP-04-24\\
April, 2004
\end{flushright}
\vspace{0.5cm}
\begin{center}
{\Large \bf
A Gravity Dual of Localized Tachyon Condensation
in Intersecting Branes \\
}
\lineskip .75em
\vskip1.5cm
\normalsize

{\large
Kazuyoshi Takahashi \footnote{
E-mail:\ \ {\tt kazuyosi@yukawa.kyoto-u.ac.jp} }
}
\vskip 2.5em

{
\it
Yukawa Institute for Theoretical Physics,
Kyoto University,\\
Kyoto 606-8502, Japan
}
\vskip 1em
\vskip 2em

{\bf Abstract}

\end{center}

The method of probe brane is the powerful
one to obtain the effective action
living on the probe brane
from supergravity.
We apply this method
to the unstable brane systems,
and understand the tachyon condensation
in the context of the open/closed duality.
First, we probe
the parallel coincident branes
by the anti-brane.
In this case,
the mass squared of the string
stretched between the probe brane
and one of the coincident branes
becomes negative infinite
in the decoupling limit.
So that the dual open string field theory
is difficult to understand.
Next, we probe parallel coincident branes by
a brane intersecting with an angle.
In this case, the stretched strings
have the tachyonic modes localized
near the intersecting point,
and by taking the appropriate limit
for the intersection angle,
we can leave mass squared of this modes
negative finite
in the decoupling limit.
Then we can obtain the information
about the localized tachyon condensation
from the probe brane action obtained
using supergravity.

\end{titlepage}

\baselineskip=0.7cm


\section{Introduction}

The importance of unstable brane systems
has been recognized for long times
\cite{Sen}.
One of the most successful utilization
of the unstable brane systems
was the tachyon condensation \cite{Sen1},
which have given the many informations
to understand the time-dependent dynamics
of the decay of unstable brane system,
and the tachyon condensation
has been investigated
using many powerful methods,
for example,
string field theory, effective field theory,
the boundary CFT (or rolling tachyon),
K-theory,
and so on.

The open/closed duality also
has been well known
as the powerful method and concept
in string theory.
And so it is hoped
that the open/closed duality
is applied to the unstable brane system
involving tachyon condensation.
In this paper,
we try to analyze some unstable branes
and tachyon condensation
using the probe brane method
which is conjectured from the open/closed duality.

The open/closed duality
is based on the following ideas.
It is well known that the D-branes
admit a perturbative description
in terms of the open strings
which are attached to them.
On the other hand, the D-branes
is thought to be a soliton
solution of supergravity
(or closed string theory).
Using these two different descriptions,
we can obtain the information of field theory
on branes from the supergravity.
The AdS/CFT correspondence
is the most successful example
\cite{Maldacena1,Klebanov}.

The method of probe brane
is the important one
based on the open/closed duality
\cite{Maldacena,Maldacena1,Imeroni}
\footnote{The original references are contained
within \cite{Maldacena}.}.
We consider a probe brane
which moves close to
a large number of coincident branes.
Then we can obtain the effective action
living on the probe brane
using supergravity.
This method has been confirmed
to be valid for
some examples not limited to the extremal cases.
In this paper,
we show that the probe brane method
is useful
in understanding the unstable brane systems
and tachyon condensation.

At first, we probe the brane solution
by the anti-brane.
In this case,
we show that the mass squared of the stretched strings
become negative infinite
in the decoupling limit ($\alpha' \rightarrow 0$),
so that the dual open string field theory
is not well defined.

Next, we probe parallel coincident branes by
a brane intersecting with an angle.
The energy of the open strings stretched between
probe brane and the one of the rest coincident branes
are minimized by being confined to the intersection point,
and these stretched strings
involve tachyon modes localized
near the intersecting point
when the probe brane is close enough
to the coincident branes.
Interestingly,
in the contrast
to the case of the anti-brane probing,
the mass squared of the localized tachyons
leaves negative finite
in the decoupling limit
by taking an appropriate limit
for the intersection angle.
Then the effective action
obtained from supergravity is nontrivial
and involves the information
about localized tachyon condensation.
We analyze
the localized tachyon condensation
from the supergravity.

In the intersecting branes at angles,
it is known that the recombination of branes
occurs
\cite{Polchinski1,Morosov,Hashimoto}
\footnote{The recombination
of branes is well known to
be an intriguing realization
of Higgs phenomena in Standard model
and so this process is also interesting
phenomenologically.
For example
See \cite{Koko}
and their reference for recent progress.},
and this is interpreted
as (local) tachyon condensation recently
\cite{Hashimoto}.
Our results from probe brane
is not incompatible with the recombination process,
and this also will be discussed.

There are the another interesting approaches
toward the open/closed duality
in unstable brane systems recently.
For example,
the open/closed duality at tree level
in the rolling tachyon process \cite{Sen2}
which insists that the tachyon matter
in open string picture
can be identified as the collective
closed strings \cite{Sen3,Sen4,Yi1},
the observation
that in the matrix model of the non-critical
string theory the matrix eigenvalue can be identified
as the open string tachyon on an unstable
$D0$-brane \cite{Verlinde,Seiberg1,Seiberg2,
Takayanagi},
Schwarzschild black brane
as the unstable brane systems \cite{Danielsson,
Guijosa,Peet,Bergman,Kalyana},
imaginary branes as the closed string states
\cite{Rastelli},
time dependent holography
in S-branes \cite{Strominger2},
the description of D-branes
in the closed string field theory
\cite{Matsuo,Matsuura}, and so on.
We believe that the method of the probe brane
together with these works
will give the some clues
to the open/closed duality
in unstable brane system.

The paper is organized as follows.
In section \ref{probe},
we review the method of probe brane
and open/closed duality
(gauge/gravity correspondence) briefly.
Next, we probe
the parallel coincident branes
by the anti-brane
and discuss about the dual field theory.
In section \ref{inter},
we probe parallel coincident branes by
a brane intersecting with an angle
and analyze the localized tachyon condensation
from the effective action obtained
using supergravity.
The relation with the recombination also
is discussed in this section.
The section \ref{conclusion}
is devoted to conclusion.

Finally we comment on
the interpretation of the probe brane
action obtained in this paper briefly.
In the anti-brane
and intersecting brane probing,
the stretched strings
include the tachyonic modes.
So that the energy scale in the action
of the probe brane is not low
compared to the masses
of the all stretched strings.
Therefore the action of the probe brane
can not be regarded as the effective
action obtained
by integrating out
the all fields whose energy scale
is low compared to the one of the stretched strings,
as is usually recognized in the probe brane
\cite{Maldacena1}.
Alternatively, the probe brane action
should be regarded as the action
which is obtained by integrating out only the fields
generated by the stretched string.
As we will see in section \ref{inter},
the energy scale of the fields on the probe brane
is proposed to be equal to the mass scale of the tachyon
field around the true vacuum of the tachyon potential.

\section{Brane and Anti-Brane Probing
\label{probe}}

The probing brane method is powerful one
for investigating the dynamics of
the large $N$ gauge theory
and this gave the important clues for
the AdS/CFT correspondence
\cite{Maldacena1,Klebanov}.
In section \ref{probing},
we review the probing brane method
and gauge/gravity correspondence briefly
following \cite{Maldacena,Maldacena1,Imeroni}.
In section \ref{anti}, we replace probe brane
with anti-brane
and discuss about the dual open string field theory
and point out the some problems.
In this paper, we consider $D3$-brane mainly.

\subsection{Brane Probe and Gauge/Gravity Correspondence
\label{probing}}

In this section, we review the method of probe brane
following \cite{Maldacena,Maldacena1,Imeroni}.
We consider $N+1$ $Dp$-branes, $N$ of which sit at the
origin ($r=0$) and the last is the probe brane
which sits at a distance $r$.
The world-volume action at low energies is
$p+1$-dimensional $U(N+1)$ Yang-Mills theory
broken down to $U(N) \times U(1)$ by the
expectation value of an adjoint scalar
which represents the distance between the probe
and the rest of branes.
The strings stretched between the probe brane
and one of the rest $N$ branes
are massive, with a mass $m=r/(2\pi\alpha')$.

We want to ignore the excited string states,
which is performed by letting $\alpha' \rightarrow 0$.
In order for the energies of stretched strings
to remain finite in this limit, we need
to approach the probe brane to
origin in such a way that
$\hat{X}^{I} \equiv \frac{X^I}{\alpha'}=\rm{fixed}$,
where $X^I$ ($I=p+1, \cdots, 9$)
is the transverse position
of the probe $Dp$-brane.
And, at same time, we need to take gauge coupling constant finite.
These limits for $Dp$-brane
are well known in the context
of AdS/CFT and given in summary by
\begin{eqnarray}
\alpha' &\rightarrow& 0
\nonumber \\
\gym^2&=&2(2\pi)^{p-2}g_{s}\alpha'^{\frac{p-3}{2}}
=\rm{fixed} \ ,
\label{decouple}
\end{eqnarray}
and near-horizon limit
\begin{eqnarray}
\hat{X}^{I} &\equiv& \frac{X^I}{\alpha'}=\rm{fixed}
\label{decouple2} \ ,
\end{eqnarray}
where $g_{s}$ is string coupling.

On the other hand, D-branes are well described
as a classical solution of the low-energy supergravity
equations of motion in the region
\begin{eqnarray}
g_{s} \rightarrow 0, \ \ N \rightarrow \infty
\nonumber \\
g_{s}N=\rm{fixed}  \gg 1,
\label{sugra}
\end{eqnarray}

We consider the $D3$-brane case mainly.
In the region (\ref{sugra}), the $N$ $D3$-branes
is well described by the
corresponding supergravity solution
\footnote{See the review \cite{Peet2} for example.}as
\begin{eqnarray}
ds^2&=&H^{-1/2}\eta_{\alpha\beta}dx^{\alpha}dx^{\beta}
+H^{1/2}\delta_{ij}dx^{i}dx^{j},
\nonumber \\
e^{\Phi}&=&1,
\nonumber \\
F_5&=&dH^{-1}\wedge dx^0\wedge\cdots
\wedge dx^3
+(dH^{-1}dx^0\wedge\cdots\wedge dx^3)^{\star},
\label{D3}
\end{eqnarray}
where $\alpha,\beta=0,\cdots 3$,
$i,j=4 \cdots 9$ and
$F_5$ is R-R five-form field strength.
The warp factor $H$ is given by
\begin{equation}
H(r)=1+\frac{4\pi g_{s}N \alpha'^2}{r^4}
\label{H} \ .
\end{equation}
In the decoupling limit (\ref{decouple}) and (\ref{decouple2}),
\begin{equation}
H(\hat{r})=\frac{1}{(\alpha')^2}\frac{4\pi g_{s}N}
{\hat{r}^4}
\end{equation}
where $\hat{r}^2=(\hat{X}^I)^2$,
and we used the $\hat{X}^I$,
which is finite in the limit (\ref{decouple}),
instead of the $X^I$.

In the parameter region (\ref{sugra}),
the $N$ $D3$-branes at origin ($r=0$) is
represented by the brane solution (\ref{D3}),
and so the probe $D3$-brane
takes the interaction from curved background (\ref{D3}).
In the gauge field side,
this interaction
comes from integrating out the massive fields
generated by the open string
which is stretched between probe brane
and one of the rest branes.
Thus, after integrating out these massive fields,
the world-volume action on the probe brane
should be equal to the Born-Infeld action
in curved background (\ref{D3}).
This suggestion is known as the open/closed duality
(gauge/gravity correspondence)
and has played an important roles
also in AdS/CFT correspondence
\cite{Maldacena1,Klebanov}.

The action of a probe $D3$-brane
in a general supergravity background is given,
for turning off the gauge field and Kalb-Ramond field, by
\begin{eqnarray}
S=-\frac{1}{g_s (2\pi)^3\alpha'^2}
\int  d^4 x \{e^{-\Phi}
\sqrt{-{\rm det} \ {\bf P}(G_{\mu \nu})}
+{\bf P} (C_{4}) \}
\label{Born}
\end{eqnarray}
where ${\bf P}$ denotes
pull-back to the world-volume of bulk fields
and $C_{4}$ is
the R-R $4$-form potential.
In the $N$ $D3$-branes background,
we obtain the action on probe brane
at leading order of the kinetic terms as
\begin{equation}
S=-\frac{1}{4 \pi^2\gym^2} \int dx^4
\del_{\mu} \hat{X}^I \del^{\mu} \hat{X}^I
 \ ,
\label{parallel}
\end{equation}
where $\mu=0,\cdots,3$ is the world-volume direction.

It is confirmed
in some examples not limited
to extremal cases
that results obtained from supergravity
match with the results which is obtained
by integrating out the stretched strings directly
in the field theory side
\footnote{For example, see the reference
in \cite{Maldacena}}.

\subsection{Anti-Brane Probing Brane Solution
\label{anti}}

In this section, we consider $N$ $D3$-branes
and one $\ol{D3}$-brane,
where $N$ $D3$-branes sit at the origin ($r=0$)
and the $\ol{D3}$-brane sits at a distance $r$ as a probe brane
\footnote{This system
was discussed originally in \cite{Rabadan},
but the relation with open/closed duality
was not investigated there.}.
We investigate this system
using the method of probe brane.

First, we consider the anti-brane probe
from the supergravity.
Because the R-R charge of the $\ol{D3}$-brane
is opposite to the one of the $D3$-brane,
the world-volume action
on the $\ol{D3}$-brane probe in the background (\ref{D3})
is given by reversing the sigh of Wess-Zumino term
from the case of brane probe in section \ref{probing}.
Therefore the action on the anti-brane probe
is given at leading order of the kinetic term by
\begin{eqnarray}
S&=&-\frac{1}{4 \pi^2 \gym^2}
\int d^4 x
\left(
\partial_{\mu}\hat{X}^I\partial^{\mu}\hat{X}^I
+\frac{4 \hat{r}^4}{\gym^2 N}\right) \ .
\label{D3bar}
\end{eqnarray}

As we have reviewed in section \ref{probing},
it is expected that the probe action (\ref{D3bar})
is given by integrating out the fields
generated by the open string stretched
between $\ol{D3}$ and one of the $N$ $D3$-branes.
The only point we should
notice is that the GSO projection
for the $3$-$\ol{3}$ strings
\footnote{
We denote the open string stretched
between $\ol{D3}$ and one of the $N$ $D3$-branes
as $3$-$\ol{3}$ string,
and the one stretched
between $D3$-branes as $3$-$3$ string.}
is opposite to the one we chose
for the $3$-$3$ strings \cite{Sen}.
The mass of the lowest excitation mode of the $3$-$\ol{3}$ strings
is given by
\begin{equation}
m^2=-\frac{1}{2\alpha'}
+\left(\frac{\hat{r}}{2\pi} \right)^2 \ .
\label{tachyon}
\end{equation}
Interestingly, the lowest mode of $3$-$\ol{3}$ strings
is tachyonic and
it is expected that the anti-brane probe action (\ref{D3bar})
is the effective action on $\ol{D3}$-brane
after integrating out these tachyon fields.
However this anticipation is naive.
We need to impose the decoupling limit (\ref{decouple})
and (\ref{decouple2})
to ignore excited string states.
In this limit,
the mass squared (\ref{tachyon})
becomes negative infinite,
so that the behavior of these fields
is not well defined.
This implies that the dual open string picture
in the $\ol{D3}$-brane probing is unclear
or invalid even though the probe action from supergravity
is well defined.

In next section,
we propose the brane configuration
where the fields integrated out
include tachyonic modes and these mass squared
leaves finite
in the decoupling limit (\ref{decouple}).
This brane configuration
is the intersecting branes at angles,
which has been discussed often
in the context of tachyon condensation recently
\cite{Hashimoto,Nagaoka1,Hashimoto2,Nagaoka2}.

\section{Intersecting Branes
and Open/Closed Duality
\label{inter}}

In this section,
we probe parallel coincident branes by
a brane intersecting with an intersection angle.
As noted in section \ref{probing},
the probe brane action obtained from supergravity
should be equal to the effective action
after integrating out
the fields generated by the stretched string.
In this case,
the stretched string
include tachyon modes localized
near the intersecting point.
Interestingly, by taking appropriate limit
for the intersection angle,
the mass squared of this tachyonic mode
leaves finite
in the decoupling limit (\ref{decouple})
in contrast to the case of the anti-brane probing.
Using the probe brane action
obtained from supergravity,
we can analyze the condensation of the localized tachyon.

In section \ref{open} and \ref{closed},
we analyze this system from field theory side
(open string side) and supergravity side
(closed string side) respectively.
In section \ref{condensation},
we discuss about the localized tachyon condensation
using the probe brane action obtained from supergravity.
In section \ref{recom},
we comment on the relation with the recombination process
in the intersecting branes
\cite{Polchinski1,Morosov,Hashimoto}.

\subsection{Intersecting Brane Probe : Field Theory Side
\label{open}}

\begin{figure}[htb]
\begin{center}
\unitlength=.4mm
\begin{picture}(200,140)(0,0)
\epsfxsize=8cm
\put(0,0){\epsfbox{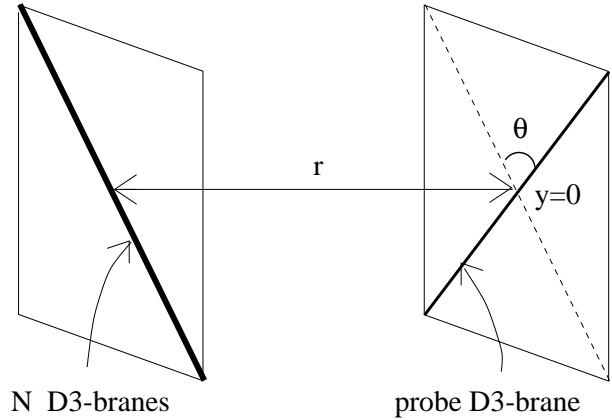}}
\end{picture}
\end{center}
\caption{
We probe $N$ parallel $D3$-branes by
an brane
which has an intersection angle $\theta$
in the $(X^3,X^4)$-plane,
and we set the intersecting point
to $y \equiv X^3 =0$.
}
\label{figure1}
\end{figure}

In this section, we probe $N$ parallel coincident
$D3$-branes by
an brane which has an intersection angle $\theta$.
The $N$ parallel $D3$-branes
are parameterized by
world-volume coordinates $X^0,X^1,X^2,X^3$,
and sit at the origin of the transverse spaces
($X^4, \cdots, X^9=0$).
The probe brane is also parameterized by
the $X^0,X^1,X^2,X^3$ coordinates,
but has intersection angle $\theta$
in the $(X^3,X^4)$-plane,
so that $X^4$ depends on $X^3 \equiv y$ coordinate as
\footnote{
We treat $X^{4}$ as the fixed background.
The fluctuation of $X^{4}$
will be given later on.}
\begin{eqnarray}
X^{4}&=&qy,
\nonumber \\
q& \equiv &\rm{tan} \, \theta \ .
\label{con}
\end{eqnarray}
The transverse distance
between probe brane and the one of rest branes
at the intersecting point ($y=0$)
is denoted as $r$, or
$r \equiv \sqrt{\sum_{I=5}^{9}(X^I)^2}
\ \Big|_{y=0}$.
See the figure \ref{figure1}
for this brane configuration.
In this case,
the energy of the open string stretched between
probe $D3$-brane and the rest of $N$ $D3$-branes
is minimized by being confined
to the intersection point ($y=0$).
It is known from world-sheet analysis
\cite{Douglas,Arfaei}
that the mass squared of this stretched open string is given by
\begin{equation}
m^2=\left( n-\frac{1}{2} \right)
\frac{\theta}{\pi \alpha'}
+\left(\frac{r}{2\pi\alpha'}\right)^2
\label{mass1}
\end{equation}
where $n$ is a integer ($n \geq 0$)\footnote{
Of course, $\theta=0$ case corresponds
to the parallel D-branes case in section \ref{probing}
and $\theta=\pi$ case corresponds to the case
in section \ref{anti},
and they give the same mass relation with (\ref{mass1})
respectively.}.
Therefore, when $r$ is small enough,
the lowest mode of the stretched strings are tachyonic
\footnote{See \cite{Ohta} for the related world-sheet analysis
of intersecting branes.}.

In contrast to the case in section \ref{anti},
in decoupling limit (\ref{decouple}),
the mass squared of the tachyon mode
is not always negative infinite,
and can be finite by taking the appropriate limit
for $\theta$ and $X$ as
\begin{eqnarray}
\hat{\theta} &\equiv& \frac{\theta}{\alpha'}={\rm fixed} \ ,
\nonumber \\
\hat{X}^I &\equiv& \frac{X^I}{\alpha'}={\rm fixed}.
\label{limit}
\end{eqnarray}
where $I=5,\cdots,9$. Then the mass squared
(\ref{mass1}) is given by
\begin{equation}
m^2=\left(n-\frac{1}{2} \right)
\frac{\hat{\theta}}{\pi}
+\left(\frac{\hat{r}}{2\pi} \right)^2 \ ,
\label{mass}
\end{equation}
where $\hat{r} \equiv r/ \alpha'$.
For $\hat{r}=0$,
the width of localized tachyons $\delta y$ is
given \cite{Morosov,Hashimoto} by
\footnote{
This quantity has not been calculated
for the case of $\hat{r} \neq 0$.}
\begin{equation}
\delta y \sim \sqrt{\hat{\theta}} \ .
\end{equation}

If we take another limit for the intersecting angle $\theta$ as
\begin{equation}
\tilde{\theta} \equiv
\frac{\theta}{(\alpha')^K}={\rm fixed} \ ,
\label{anotherlimit}
\end{equation}
the $\theta$ part does not contribute
to the mass squared (\ref{mass1}) for $K > 1$,
and gives negative infinite mass
to the mass squared for $K < 1$.
These results
are reflected to the probe action
obtained from supergravity,
as we will discuss
in the next section.

As in section \ref{probing},
it is anticipated from the open/closed duality
that the Born-Infeld action on the probe brane
in curved background (\ref{D3})
should be equal to the action on probe brane
after integrating out the fields
generated by the stretched strings.
Interestingly, in this case,
the lowest modes of the fields integrated out
are tachyonic
when the distance $\hat{r}$ is small enough.
Integrating out
the tachyonic modes do not always imply
that the effective action is not well-defined,
as well known in the case of the brane-antibrane system
\cite{Sen1}
\footnote{In early days,
some calculations to suggest that
the tachyon condense and lead to the stable vacuum
was performed
\cite{Halpern}.}.

As we have denoted in the introduction,
we can not regard the probe brane action
as the effective action
obtained
by integrating out
the all fields whose energy scale
is low compared to the one of the stretched strings,
as is usually recognized in the probe brane
\cite{Maldacena1}.
This occurs due to the tachyonic modes,
whose mass is not defined,
so that the energy scale in the action
of the probe brane is not low
compared to the masses
of the all stretched strings.
Therefore the action of the probe brane
should be regarded as the action
which is obtained by integrating out only the fields
generated by the stretched string
rather than the effective action.
However the energy scale
of the fields on the probe brane
is unclear in this picture.
We propose that this energy scale
is equal to the mass scale of tachyon field
around the true vacuum of the tachyon potential.
This proposal is based on the ideas
that the probe brane action
is equivalent to the effective action
which is defined
at the true vacuum of the tachyon potential
as we will see in section \ref{condensation}.

In the next section,
we derive the probe brane action
from supergravity
and show how the limit (\ref{limit})
or (\ref{anotherlimit}) are reflected
to the probe brane action.

\subsection{Intersecting Brane Probe : Supergravity Side
\label{closed}}

We derive probe brane action
from the supergravity
as in section \ref{probing}.
The induced metric on the probe brane
is given by
\begin{eqnarray}
{\bf P}(G_{\mu \nu})&=&H^{-\frac{1}{2}}
\left( \eta_{\mu \nu}
+\frac{\del X^I}{\del x^{\mu}}
\frac{\del X^I}{\del x^{\nu}}H \right)
\ \ {\rm for \ \mu, \nu \neq 3}
\nonumber \\
{\bf P}(G_{33})&=&H^{-\frac{1}{2}}
\left\{ 1+\left(q^2+\left(\frac{\del X^I}{\del y}\right)^2
\right)H \right\} \ ,
\label{projection}
\end{eqnarray}
where we denoted $X^I$ as
$X^I=(X^5, \cdots ,X^9)$
and $H$ is given by (\ref{H}).

The action on the probe brane
is given from (\ref{Born}) and (\ref{projection}) by
\begin{eqnarray}
S&=&-\frac{1}{g_s (2 \pi )^3 \alpha'^2}
\int dx^4 
H^{-1}\sqrt{\prod_{i=0}^{3}
\left( \eta_{ii}+\left(\frac{\del X^I}{\del x^i}\right)^2H \right)
\left\{ 1+\left(q^2+\left(\frac{\del X^I}{\del y}\right)^2
\right)H \right\}}
\nonumber \\
&+& \frac{1}{g_s (2 \pi )^3 \alpha'^2}
\int dx^4 (H^{-1}-1) \ .
\label{probeaction}
\end{eqnarray}
Here we ignored the contribution
from the non-diagonal part of ${\rm P}(G_{\mu \nu})$
in the determinant
to consider only the leading order
of the kinetic terms afterward.

First, we take the limit (\ref{decouple})
and (\ref{limit}) in the action (\ref{probeaction}).
In the limit (\ref{decouple}) and (\ref{limit}),
$H$ is represented by
\begin{eqnarray}
H&=&1+\frac{4\pi g_{s}N \alpha'^2}
{\left( \left(qy \right)^2+
\left( X^I \right)^2 \right)^2}
\nonumber \\
&=&\frac{1}{\alpha'^2}
\left(\alpha'^2
+\frac{\gym^2 N}
{\left( (\hat{\theta}y)^2+
(\hat{X}^I)^2 \right)^2}
\right)
\end{eqnarray}
where we used the relation
$q \equiv {\rm tan} \theta \rightarrow \theta$
in the limit (\ref{limit}).
Therefore it is convenient to denote $H$ as
\begin{equation}
H=\frac{\hat{H}}{\alpha'^2} \ ,
\label{Hrelation}
\end{equation}
where $\hat{H}$ is given by
\begin{eqnarray}
\hat{H} &\equiv&
\frac{\gym^2 N}{\hat{r}^4}
\nonumber \\
\hat{r} &\equiv&
\sqrt{ \left(\hat{\theta}y \right)^2+
\left( \hat{X}^I \right)^2 } \ .
\label{Hhat}
\end{eqnarray}
We note that $\hat{H}$ leaves finite in the limit
(\ref{decouple}) and (\ref{limit}).
Using $\hat{H}$ the probe action is given by
\begin{eqnarray}
S&=&-\frac{1}{2 \pi^2 \gym^2}
\int dx^4 
\hat{H}^{-1}\sqrt{\prod_{i=0}^{3}
\left( \eta_{ii}+\left(\frac{\del \hat{X}^I}
{\del x^i}\right)^2\hat{H} \right)
\left\{ 1+\left(\hat{\theta}^2
+\left(\frac{\del \hat{X}^I}{\del y}\right)^2
\right)\hat{H} \right\}}
\nonumber \\
&+& \frac{1}{2 \pi^2 \gym^2}
\int dx^4 \hat{H}^{-1}
\label{probe1}
\end{eqnarray}
where we denoted as $\hat{X}^I \equiv X^I/ \alpha'$
and used $\gym$ in (\ref{decouple}).
This action is finite and well-defined
in the limit (\ref{decouple})
and (\ref{limit}),
which has a physical implication
in the contrast of the anti-brane case
in section \ref{anti}.

Now we consider
only the leading order terms
in the kinetic terms to analyze the dynamics
of the probe brane.
Then the probe brane action is given by
\begin{eqnarray}
S&=&-\frac{1}{2 \pi^2 \gym^2} \int dx^4
\left\{\frac{\sqrt{1+\hat{\theta}^2\hat{H}}}{2}
\left(\frac{\del \hat{X}^I}{\del x^{\alpha}}\right)^2
+\frac{1}{2\sqrt{1+\hat{\theta}^2\hat{H}}}
\left(\frac{\del \hat{X}^I}{\del y}\right)^2
\right\}
\nonumber \\
&-&\frac{1}{2 \pi^2 \gym^2} \int dx^4
\left(\hat{H}^{-1}\sqrt{1+\hat{\theta}^2\hat{H}}
-\hat{H}^{-1} \right)
\label{kinetic}
\end{eqnarray}
where $x^{\alpha}=(x^0,x^1,x^2)$.
Therefore the coefficients of kinetic terms
depend on $\hat{\theta}$ and $\hat{H}$
in contrast with the case of the parallel branes
(\ref{parallel}).

For $\hat{X}^I \gg 1$ and $y \gg 1$,
because $\hat{H} \simeq 0$
from (\ref{Hhat}),
the action (\ref{kinetic})
is equal to the parallel brane case
(\ref{parallel}).
On the other hand,
for $\hat{X}^I \ll 1$ and $y \ll 1$
the equation (\ref{kinetic}) involves
some physical implications,
for example, localized tachyon condensation,
which will be discussed in section \ref{condensation}.

We discuss about the another limit (\ref{anotherlimit})
for $\theta$.
We see how the limit
for the intersection angle
is reflected to the probe brane action
in supergravity.
We represent the probe action (\ref{probeaction})
using $\tilde{\theta}$ in (\ref{anotherlimit})
and $\hat{X}$.
For $K > 1$,
it is easy to show that
the probe action is equal to the action of probe brane in the parallel
$D3$-branes case in section \ref{probing}.
This is expected result from open string picture,
because the mass of the fields integrated out
is equal to the case in section \ref{probing}
in decoupling limit.

The case of $K < 1$ is somewhat nontrivial.
In the field theory side,
the mass of the field integrated out
is negative infinite,
so that the action on the probe brane
is difficult to understand or invalid.
Actually, the probe brane action
obtained from supergravity is compatible
with the result from the open string picture.
For $K < 1$, it is easy to confirm
that the probe action
(\ref{probeaction}) is order $(\alpha')^{2(K-1)}$,
so that the action becomes infinite
in the decoupling limit (\ref{decouple}).
Therefore this action is not well defined.
This compatibility with open string picture
is one of the evidence for the open/closed duality
in the intersecting brane system.

Finally in this section,
we consider the case
where the fluctuation of $X^4$
exists and be given by
\begin{equation}
X^{4}=qy+\tilde{X}^4(x,y) \ ,
\label{con2}
\end{equation}
where we denoted $x$ as $x=(X^0,X^1,X^2)$.
The part coming from this fluctuation
in the probe brane action
is given by
\begin{equation}
\tilde{S}=-\frac{1}{2 \pi^2 \gym^2} \int dx^4
\left\{\frac{\sqrt{1+\hat{\theta}^2\hat{H}}}{2}
\left(\frac{\del \hat{X}^4}{\del x^{\alpha}}\right)^2
+\frac{1}{2\sqrt{1+\hat{\theta}^2\hat{H}}}
\left(2 \hat{\theta}\frac{\del \hat{X}^{4}}{\del y}
+\left(\frac{\del \hat{X}^4}{\del y}\right)^2
\right)
\right\} \ ,
\label{fluctuation}
\end{equation}
where we defined $\hat{X}^4$ as
$\hat{X}^{4} \equiv \tilde{X^4}/ \alpha'$.
This term is different from the kinetic term
of (\ref{kinetic}) due to the existence
of the part of first derivative of $y$.

\subsection{Localized Tachyon Condensation
\label{condensation}}

In this section,
we give a physical interpretation
to the probe brane action (\ref{kinetic}).
As we will discuss,
this action involves the information
about the (localized) tachyon condensation.

First, we remember the tachyon condensation
in a brane-antibrane system
\cite{Sen1,Sen}.
A system of coincident
brane-antibrane has tachyonic modes
generalized by the open string stretched
between brane and anti-brane,
and Sen conjectured
that there is no degree of freedom
of the open string around the true vacuum
of tachyon potential $V(T)$.
In this system,
if we integrate out this open string tachyon,
the partition function is dominated
at the tachyon vacuum, that is
\begin{equation}
\int [DT]\exp (-S[T])
\sim \exp(-S[T_0])
\end{equation}
where $T=T_0$ is the vacuum of tachyon potential.
This implies that integrating out tachyon
gets rid of the degrees of freedom of open strings.

Same ideas can be applied
to the case of the intersecting branes.
As we have denoted in section \ref{open},
for the probe brane being enough close to the origin,
there are localized tachyon modes generated by
the open strings stretched between probe brane
and one of the rest branes.
Of course, the potential
of the localized tachyon is not bounded
at tree level,
but by integrating out the localized massive mode
($n > 0$) in (\ref{mass})
this potential will be bounded.
Then if we integrate out these tachyons localized
at the intersecting point ($y=0$),
we expect that the partition function
dominates at the vacuum of the potential
and that there are no degrees of freedom
of open strings near the intersecting point
($y=0$) for $\hat{X}^I=0$.
These can be seen
by the probe brane action obtained
from supergravity.

As we noted in section \ref{open},
it is expected that
the probe brane action (\ref{kinetic})
obtained from supergravity
should be equal to the action on probe brane
after integrating out the fields
generated by the open string
stretched between probe brane and one
of the rest branes, which include
tachyonic modes
for probe brane enough close to origin.
Therefore, in this case, the action (\ref{kinetic})
should not have degrees of freedom of open string
near the intersecting point,
namely $\hat{X}^I \ll 1$
and $y \ll 1$.

Actually we can see the absence
of the degrees of freedom of open strings
in the probe brane action (\ref{kinetic}).
At first, we consider about the potential part,
which is given by
\begin{equation}
V_{\rm pot}=\frac{1}{2 \pi^2 \gym^2}
\left(\hat{H}^{-1}\sqrt{1+\hat{\theta}^2\hat{H}}
-\hat{H}^{-1} \right) \ .
\end{equation}
Then, because
$V_{\rm pot} \rightarrow \hat{\theta}/(2 \pi^2 \gym^2)$
for $\hat{X}^I \rightarrow 0$ and $y \rightarrow 0$,
the potential part is almost a constant
in the region $\hat{X}^I \ll 1$ and $y \ll 1$.

Because the potential part is almost constant,
the kinetic terms (\ref{kin}) should involve
the information about the dynamics of the probe brane.
The part of kinetic terms
in (\ref{kinetic})
is given by
\begin{equation}
S_{\rm kin}=-\frac{1}{2 \pi^2 \gym^2} \int dx^4
\left\{\frac{\sqrt{1+\hat{\theta}^2\hat{H}}}{2}
\left(\frac{\del \hat{X}^I}{\del x^{\alpha}}\right)^2
+\frac{1}{2\sqrt{1+\hat{\theta}^2\hat{H}}}
\left(\frac{\del \hat{X}^I}{\del y}\right)^2
\right\} \ ,
\label{kin}
\end{equation}
where $x^{\alpha}=(x^0,x^1,x^2)$, $y=x^3$
and $I=5, \cdots ,9$.
Because $\hat{H} \gg 1$ in this case,
the coefficient of $\del \hat{X}^I/ \del x^{\alpha}$
is large. This implies
that the $x^{\alpha}$ dependence of $\hat{X}^I$ is
classical and does not have quantum fluctuations
\footnote{This is similar
to the case of zero temperature in thermal system.}.
Therefore for $\hat{X}^I \ll 1$ and $y \ll 1$,
the degrees of freedom of $\hat{X}^I$
along the directions of $x^{\alpha}$
have been lost,
so that $\hat{X}^I$ fluctuate following only $y$.

On the other hand,
for $\hat{X}^I \ll 1$ and $y \ll 1$,
the coefficient of the $\del \hat{X}^I/ \del y$
becomes very small.
This is similar to the absence of open string
in the tachyon condensation of non-BPS brane.

To see this similarity,
we remember the tachyon condensation
in the non-BPS branes.
In the case of the non-BPS $p$-branes
\footnote{The $p$ is odd for Type IIA
string theory and even
for Type IIB string theory.},
the world-volume action in flat space
is given \cite{Sen5,Sen2} by
\begin{equation}
S=-\int d^{p+1} x V(T)
\sqrt{-\det \left( \eta_{\mu \nu}
+\del_{\mu}X^i\del_{\nu}X^i
+\del_{\mu}T\del_{\nu}T 
+2\pi \alpha' F_{\mu\nu} \right)} \ ,
\end{equation}
where  $V(T)$ is tachyon potential,
$X^i$ ($i=p+1,\cdots, 9$)
is the transverse scalar,
and we have omitted the Wess-Zumino.
Because $V(T) \rightarrow 0$ for $T \rightarrow T_{0}$,
the coefficient of the $\del X^I/ \del x^{\mu}$
approach zero in the process of tachyon condensation.
And so the non-BPS brane action
after the tachyon condensation has the similar behavior
with the kinetic term $\del \hat{X}^I/ \del y$
in intersecting branes.
This implies that the degrees of freedom of open string
along the direction $y$ in intersecting branes
are absent in the sense similar
with the case of the non-BPS brane
\footnote{In the case of non-BPS brane,
it is proposed that the degrees of freedom
of the open string don't die in tachyon condensation,
but leave
as the ones of the string fluid
\cite{Yi2},
which is interpreted as the collective closed strings
recently \cite{Sen3,Sen4,Yi1}.}.

In summary,
the probe brane action (\ref{kin})
seems to imply
that the degrees of freedom of open string near
the intersecting point
become absent due to the localized tachyon condensation
\footnote{The kinetic term of the fluctuation $\tilde{X}^4$
(\ref{fluctuation}) is not similar to the non-BPS case
due to the part of the first derivative of $y$.
This result has not been understood physically.}.
However, about how degrees of freedom become absent,
the case of the $x^{\alpha}$ dependence
and one of the $y$ dependence of $\hat{X}^I$
differ from each other.
This physical implication has been unclear yet.

Finally in this section,
we point out the naive point
in this subsection.
The mass relation (\ref{mass})
is the one at tree level.
Therefore, the tachyonic modes
at tree level can become
massless or massive
due to the loop correction,
so that the localized tachyon condensation
could not occur.
But, as we noted before, it seems
that the probe brane action (\ref{kin})
shows the absence of degrees of freedom
of open strings.
This argument is not rigorous one,
but we believe that it is a evidence of the occurrence
of the localized tachyon condensation
\footnote{In the paper \cite{Park}
the supergravity solutions
dual to microstates of the D1-D3-D5 system
with nonzero B field moduli is given.
Because the angles of branes are
T-dual to the B field,
the relation with our papers and the paper
\cite{Park} is interesting to investigate.}.

\subsection{On Recombination of Intersecting Branes
\label{recom}}

\begin{figure}[htb]
\begin{center}
\unitlength=.4mm
\begin{picture}(370,100)(0,0)
\epsfxsize=15cm
\put(0,0){\epsfbox{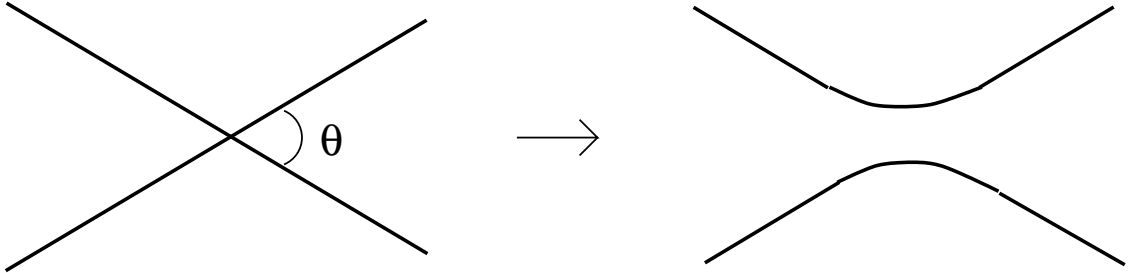}}
\end{picture}
\end{center}
\caption{The intersecting branes are
recombined due to the tachyon mode
localized at the intersecting point.
}
\label{recombination}
\end{figure}

In the intersecting branes at angles,
it is known that the recombination of branes
occurs
\cite{Polchinski1}.
The recombination
can be analyzed by the effective field theory
on the branes \cite{Morosov,Hashimoto},
and it was interpreted
as local tachyon condensation
recently in \cite{Hashimoto}.
See figure \ref{recombination}.

In the same way,
in our cases, where $N$ branes are parallel
to each other and only a probe brane
is intersected at an angle,
it is expected that the recombination
of the probe brane with one of the rest branes
will occur.
However, the probe brane action does not
show the recombination of branes.
At first sight, this seems
to imply the incompatibility
between the recombination and our results
from supergravity,
but actually these two results
are compatible.

To see this,
we remember how the recombination can be
understood using effective field theory,
following \cite{Hashimoto}.
At first, it is convenient that
we consider the system
consisting of two intersecting $D3$-branes
which has an intersection angle
$\theta$ in $(X^3,X^4)$-plane,
and both branes are parameterized by
the coordinate
$x \equiv (X^0,X^1,X^2)$ and $y \equiv X^3$.
The transverse scalar field
is represented by
\begin{equation}
X^4=\left(
\begin{array}{cc}
qy & 0 \\
0 & -qy \\
\end{array}
\right)
\end{equation}
where
\begin{equation}
q \equiv
\tan(\theta / 2) \ .
\end{equation}
The non-diagonal parts
of $X^4$ exists as the fluctuation
$T(x,y)$ as
\begin{equation}
X^4=\left(
\begin{array}{cc}
qy & T(x,y) \\
T(x,y) & -qy \\
\end{array}
\right) \ .
\end{equation}
This fluctuation includes the tachyonic mode,
and now we focus on this mode.
The tachyonic mode rolls the tachyon potential,
so that this mode is time dependent
and given by
\begin{equation}
T(x,y)=C \exp\left(-\frac{qy^2}{2 \pi \alpha'} \right)
e^{\sqrt{\frac{q}{\pi \alpha'}} \,X^0} \ ,
\end{equation}
where $C$ is a constant.
The recombination
is realized by diagonalizing $X^4$,
and the diagonalized position is given by
\begin{equation}
X^4=\left(
\begin{array}{cc}
\sqrt{q^2 y^2+T(x,y)^2} & 0 \\
0 & -\sqrt{q^2 y^2+T(x,y)^2} \\
\end{array}
\right) \ .
\end{equation}

Thus, in effective field theory,
we does not integrate out the tachyon mode
and this mode evolves following the equation
of motion.
On the other hand,
in the probe brane method,
the tachyonic mode needs to be integrated out,
that is, the tachyon mode must sit
at the vacuum of the tachyon potential.
Therefore we haven't analyzed the time evolution
of branes' positions in probe brane method.
This implies that the subjects treated
in effective field theory
are not equal to the ones treated
in probing brane method,
so that these two results are compatible.

Then how the recombination
is understood in probe brane method~?
The diagonalization
of the matrix $X^4$
representing the branes' positions
needs to be taken to realize the recombination.
But this prescription is unclear
in the probe brane method,
because the non-diagonal part of the matrix $X^4$
must be integrated out.
And so, understanding recombination
from supergravity is difficult
and that is a future problem.

Finally, we emphasize that the subject of this paper
is not to study the dynamics of the intersecting branes
(recombinations and so on),
but to understand and analyze
the localized tachyon condensation
in the intersecting branes
in the context of open/closed duality
even if we treat brane configuration
as the fixed background.

\section{Conclusions
\label{conclusion}}

We have applied the probe brane method
to the unstable brane systems,
in particular intersecting branes at angles.
In this system,
we can take an appropriate
limit for the intersection angle
with the decoupling limit
$\alpha' \rightarrow 0$
to leave the mass squared of the localized tachyons
negative finite.
So that we have obtained
the probe brane action from supergravity
which involves the information
about localized tachyon condensation,
though some unclear and naive problems
have been left.
We hope that a lot of techniques
in AdS/CFT correspondence
will be applied to the intersecting branes,
or another unstable brane systems.

\section*{Acknowledgments}
\vskip2mm

I would like to thank my colleagues in Yukawa Institute for
Theoretical Physics for discussions and encouragement.
I am especially grateful to K.~Hotta, H.~Kajiura, S.~Sugimoto,
D.~Tomino for useful discussions.


\begin{thebibliography}{999}

\bibitem{Sen}
A.~Sen,
``Non-BPS states and branes in string theory,''
arXiv:hep-th/9904207.

\bibitem{Sen1}
A.~Sen,
``Stable non-BPS bound states of BPS D-branes,''
JHEP {\bf 9808}, 010 (1998)
[arXiv:hep-th/9805019]. \\
A.~Sen,
``Tachyon condensation on the brane antibrane system,''
JHEP {\bf 9808}, 012 (1998)
[arXiv:hep-th/9805170].

\bibitem{Maldacena1}
J.~M.~Maldacena,
``The large N limit of superconformal field theories and supergravity,''
Adv.\ Theor.\ Math.\ Phys.\  {\bf 2}, 231 (1998)
[Int.\ J.\ Theor.\ Phys.\  {\bf 38}, 1113 (1999)]
[arXiv:hep-th/9711200].

\bibitem{Klebanov}
S.~S.~Gubser, I.~R.~Klebanov and A.~M.~Polyakov,
``Gauge theory correlators from non-critical string theory,''
Phys.\ Lett.\ B {\bf 428}, 105 (1998)
[arXiv:hep-th/9802109].\\
E.~Witten,
``Anti-de Sitter space and holography,''
Adv.\ Theor.\ Math.\ Phys.\  {\bf 2}, 253 (1998)
[arXiv:hep-th/9802150].

\bibitem{Maldacena}
J.~M.~Maldacena,
``Branes probing black holes,''
Nucl.\ Phys.\ Proc.\ Suppl.\  {\bf 68}, 17 (1998)
[arXiv:hep-th/9709099].

\bibitem{Imeroni}
E.~Imeroni,
``The gauge/string correspondence towards realistic gauge theories,''
arXiv:hep-th/0312070.

\bibitem{Polchinski1}
J.~Polchinski,
``String Theory II,''
Cambridge University Press (1998)

\bibitem{Morosov}
A.~V.~Morosov,
``Classical decay of a non-supersymmetric configuration of two  D-branes,''
Phys.\ Lett.\ B {\bf 433}, 291 (1998)
[arXiv:hep-th/9803110].

\bibitem{Hashimoto}
K.~Hashimoto and S.~Nagaoka,
``Recombination of intersecting D-branes by local tachyon condensation,''
JHEP {\bf 0306}, 034 (2003)
[arXiv:hep-th/0303204].

\bibitem{Koko}
C.~Kokorelis,
``Exact standard model structures from intersecting D5-branes,''
Nucl.\ Phys.\ B {\bf 677}, 115 (2004)
[arXiv:hep-th/0207234]. \\
C.~Kokorelis,
``Exact standard model structures from intersecting branes,''
arXiv:hep-th/0210004.


\bibitem{Sen2}
A.~Sen,
``Rolling tachyon,''
JHEP {\bf 0204}, 048 (2002)
[arXiv:hep-th/0203211]. \\
A.~Sen,
``Tachyon matter,''
JHEP {\bf 0207}, 065 (2002)
[arXiv:hep-th/0203265]. \\
A.~Sen,
``Field theory of tachyon matter,''
Mod.\ Phys.\ Lett.\ A {\bf 17}, 1797 (2002)
[arXiv:hep-th/0204143].

\bibitem{Sen3}
A.~Sen,
``Open and closed strings from unstable D-branes,''
Phys.\ Rev.\ D {\bf 68}, 106003 (2003)
[arXiv:hep-th/0305011].

\bibitem{Sen4}
A.~Sen,
``Open-closed duality at tree level,''
Phys.\ Rev.\ Lett.\  {\bf 91}, 181601 (2003)
[arXiv:hep-th/0306137].

\bibitem{Yi1}
H.~U.~Yee and P.~Yi,
``Open/closed duality, unstable D-branes, and coarse-grained closed
strings,''
arXiv:hep-th/0402027.

\bibitem{Verlinde}
J.~McGreevy and H.~Verlinde,
``Strings from tachyons: The c = 1 matrix reloaded,''
JHEP {\bf 0312}, 054 (2003)
[arXiv:hep-th/0304224].

\bibitem{Seiberg1}
I.~R.~Klebanov, J.~Maldacena and N.~Seiberg,
``D-brane decay in two-dimensional string theory,''
JHEP {\bf 0307}, 045 (2003)
[arXiv:hep-th/0305159].

\bibitem{Seiberg2}
M.~R.~Douglas, I.~R.~Klebanov,
D.~Kutasov, J.~Maldacena, E.~Martinec and N.~Seiberg,
``A new hat for the c = 1 matrix model,''
arXiv:hep-th/0307195.

\bibitem{Takayanagi}
T.~Takayanagi and N.~Toumbas,
``A matrix model dual of type 0B string theory in two dimensions,''
JHEP {\bf 0307}, 064 (2003)
[arXiv:hep-th/0307083].

\bibitem{Danielsson}
U.~H.~Danielsson, A.~Guijosa and M.~Kruczenski,
``Brane-antibrane systems at finite temperature and the entropy of black
JHEP {\bf 0109}, 011 (2001)
[arXiv:hep-th/0106201]. \\
U.~H.~Danielsson, A.~Guijosa and M.~Kruczenski,
``Black brane entropy from brane-antibrane systems,''
Rev.\ Mex.\ Fis.\  {\bf 49S2}, 61 (2003)
[arXiv:gr-qc/0204010].

\bibitem{Guijosa}
A.~Guijosa, H.~H.~Hernandez Hernandez and H.~A.~Morales Tecotl,
``The entropy of the rotating charged black threebrane from a brane-antibrane
system,''
arXiv:hep-th/0402158.

\bibitem{Peet}
O.~Saremi and A.~W.~Peet,
``Brane-antibrane systems and the thermal life of neutral black holes,''
arXiv:hep-th/0403170.

\bibitem{Bergman}
O.~Bergman and G.~Lifschytz,
``Schwarzschild black branes from unstable D-branes,''
arXiv:hep-th/0403189.

\bibitem{Kalyana}
S.~Kalyana Rama,
``A description of Schwarzschild black holes in terms of intersecting M-branes
and antibranes,''
arXiv:hep-th/0404026.

\bibitem{Rastelli}
D.~Gaiotto, N.~Itzhaki and L.~Rastelli,
``Closed strings as imaginary D-branes,''
arXiv:hep-th/0304192.

\bibitem{Strominger2}
F.~Quevedo, G.~Tasinato and I.~Zavala,
``S-branes, negative tension branes and cosmology,''
arXiv:hep-th/0211031.\\
C.~P.~Burgess, P.~Martineau, F.~Quevedo, G.~Tasinato and I.~Zavala C.,
``Instabilities and particle production in S-brane geometries,''
JHEP {\bf 0303}, 050 (2003)
[arXiv:hep-th/0301122].\\
A.~Maloney, A.~Strominger and X.~Yin,
``S-brane thermodynamics,''
JHEP {\bf 0310}, 048 (2003)
[arXiv:hep-th/0302146].

\bibitem{Matsuo}
I.~Kishimoto, Y.~Matsuo and E.~Watanabe,
``Boundary states
as exact solutions of (vacuum) closed string field  theory,''
Phys.\ Rev.\ D {\bf 68}, 126006 (2003)
[arXiv:hep-th/0306189]. \\
I.~Kishimoto, Y.~Matsuo and E.~Watanabe,
``A universal nonlinear relation among boundary states
in closed string field theory,''
Prog.\ Theor.\ Phys.\  {\bf 111}, 433 (2004)
[arXiv:hep-th/0312122].

\bibitem{Matsuura}
T.~Asakawa, S.~Kobayashi and S.~Matsuura,
``Closed string field theory with dynamical D-brane,''
JHEP {\bf 0310}, 023 (2003)
[arXiv:hep-th/0309074].

\bibitem{Peet2}
A.~W.~Peet,
``TASI lectures on black holes in string theory,''
arXiv:hep-th/0008241.

\bibitem{Rabadan}
C.~P.~Burgess, P.~Martineau, F.~Quevedo and R.~Rabadan,
``Branonium,''
JHEP {\bf 0306}, 037 (2003)
[arXiv:hep-th/0303170].

\bibitem{Nagaoka1}
S.~Nagaoka,
``Fluctuation analysis of non-Abelian Born-Infeld action in the background
intersecting D-branes,''
Prog.\ Theor.\ Phys.\  {\bf 110}, 1219 (2004)
[arXiv:hep-th/0307232].

\bibitem{Hashimoto2}
K.~Hashimoto and W.~Taylor,
``Strings between branes,''
JHEP {\bf 0310}, 040 (2003)
[arXiv:hep-th/0307297].

\bibitem{Nagaoka2}
S.~Nagaoka,
``Higher dimensional recombination of intersecting D-branes,''
JHEP {\bf 0402}, 063 (2004)
[arXiv:hep-th/0312010].

\bibitem{Douglas}
M.~Berkooz, M.~R.~Douglas and R.~G.~Leigh,
``Branes intersecting at angles,''
Nucl.\ Phys.\ B {\bf 480}, 265 (1996)
[arXiv:hep-th/9606139].

\bibitem{Arfaei}
H.~Arfaei and M.~M.~Sheikh Jabbari,
``Different D-brane interactions,''
Phys.\ Lett.\ B {\bf 394}, 288 (1997)
[arXiv:hep-th/9608167]. \\
M.~M.~Sheikh Jabbari,
``Classification of different branes at angles,''
Phys.\ Lett.\ B {\bf 420}, 279 (1998)
[arXiv:hep-th/9710121].

\bibitem{Ohta}
N.~Ohta and P.~K.~Townsend,
``Supersymmetry of M-branes at angles,''
Phys.\ Lett.\ B {\bf 418}, 77 (1998)
[arXiv:hep-th/9710129].
T.~Kitao, N.~Ohta and J.~G.~Zhou,
``Fermionic zero mode and string creation between D4-branes at angles,''
Phys.\ Lett.\ B {\bf 428}, 68 (1998)
[arXiv:hep-th/9801135].

\bibitem{Halpern}
K.~Bardakci and M.~B.~Halpern,
``Explicit Spontaneous Breakdown In A Dual Model,''
Phys.\ Rev.\ D {\bf 10}, 4230 (1974). \\
K.~Bardakci and M.~B.~Halpern,
``Explicit Spontaneous Breakdown In A Dual Model. 2. N Point Functions,''
Nucl.\ Phys.\ B {\bf 96} (1975) 285.

\bibitem{Sen5}
A.~Sen,
``Supersymmetric world-volume action for non-BPS D-branes,''
JHEP {\bf 9910}, 008 (1999)
[arXiv:hep-th/9909062]. \\
M.~R.~Garousi,
``Tachyon couplings on non-BPS D-branes and Dirac-Born-Infeld action,''
Nucl.\ Phys.\ B {\bf 584}, 284 (2000)
[arXiv:hep-th/0003122]. \\
E.~A.~Bergshoeff, M.~de Roo, T.~C.~de Wit, E.~Eyras and S.~Panda,
``T-duality and actions for non-BPS D-branes,''
JHEP {\bf 0005}, 009 (2000)
[arXiv:hep-th/0003221]. \\
J.~Kluson,
``Proposal for non-BPS D-brane action,''
Phys.\ Rev.\ D {\bf 62}, 126003 (2000)
[arXiv:hep-th/0004106]. \\
M.~R.~Garousi,
``On-shell S-matrix and tachyonic effective actions,''
Nucl.\ Phys.\ B {\bf 647}, 117 (2002)
[arXiv:hep-th/0209068]. \\
D.~Kutasov and V.~Niarchos,
``Tachyon effective actions in open string theory,''
Nucl.\ Phys.\ B {\bf 666}, 56 (2003)
[arXiv:hep-th/0304045].

\bibitem{Yi2}
G.~W.~Gibbons, K.~Hori and P.~Yi,
``String fluid from unstable D-branes,''
Nucl.\ Phys.\ B {\bf 596}, 136 (2001)
[arXiv:hep-th/0009061]. \\
O.~K.~Kwon and P.~Yi,
``String fluid, tachyon matter, and domain walls,''
JHEP {\bf 0309}, 003 (2003)
[arXiv:hep-th/0305229].

\bibitem{Park}
O.~Lunin, S.~D.~Mathur, I.~Y.~Park and A.~Saxena,
``Tachyon condensation and 'bounce' in the D1-D5 system,''
Nucl.\ Phys.\ B {\bf 679}, 299 (2004)
[arXiv:hep-th/0304007].


\end{thebibliography}
\end{document}